\documentclass[preprint,nofootinbib,aps,superscriptaddress,eqsecnum]{revtex4-1}
\pdfoutput=1
\usepackage{float}
\usepackage{graphicx}
\usepackage{xcolor}
\usepackage{amsmath}
\usepackage{amsfonts}
\usepackage{amssymb}
\usepackage[bookmarks=true,bookmarksopen=true,
                    bookmarksnumbered=true,
                    colorlinks=true,linkcolor=black,
                    citecolor=red]{hyperref}
\usepackage{subfigure}
\usepackage{slashed}
 \usepackage{setspace} 
\usepackage{caption}
\usepackage{color}
\allowdisplaybreaks
\def\bea{\begin{eqnarray}}
\def\eea{\end{eqnarray}}
\def\be{\begin{equation}}
\def\ee{\end{equation}}

\begin{document}
\title{Flavor changing Flavon decay $\phi\to tc$ ($\phi=H_F,\,A_F$) at the High Luminosity Large Hadron Collider}

\author{M. A. Arroyo-Ure\~na}
\email{marcofis@yahoo.com.mx}
\affiliation{Centro Interdisciplinario de Investigaci\'on y Ense\~nanza de la Ciencia (CIIEC),\\
  Benem\'erita Universidad Aut\'onoma de Puebla,\\
 C.P. 72570, Puebla, Pue., Mexico.}

\author{A. Fern\'andez-T\'ellez}
\email{afernand@fcfm.buap.mx}
\affiliation{ Facultad de Ciencias F\'isico-Matem\'aticas,\\
Benem\'erita Universidad Aut\'onoma de Puebla,\\
C.P. 72570, Puebla, Pue., Mexico.}


\author{G. Tavares-Velasco}
\email{gtv@fcfm.buap.mx}
\affiliation{ Facultad de Ciencias F\'isico-Matem\'aticas,\\
Benem\'erita Universidad Aut\'onoma de Puebla,\\
C.P. 72570, Puebla, Pue., Mexico.}

\begin{abstract}
We present a study of the flavor changing decays $\phi\to tc$ ($\phi=H_F,\,A_F$) of the $CP$-even and $CP$-odd scalar flavons at the large hadron collider and its next stage, the high-luminosity large hadron collider. The theoretical framework is an extension of the standard model that incorporates an extra complex singlet and  invokes the Froggatt-Nielsen mechanism with an Abelian flavor symmetry. The projected exclusion and discovery regions in terms of the model parameters are reported. We find that $A_F$ could be detected at the LHC by considering a reasonable scenario of the model parameter space. As far as $H_F$ is concerned, we also found promising results that could be verified experimentally at the high-luminosity LHC.



\end{abstract}


\maketitle


\section{Introduction}
It is well known that the standard model (SM) has been successful in predicting results experimentally tested to a high accuracy, culminating with  the recent discovery of a new scalar boson compatible with the SM Higgs boson \cite{ATLAS:2012yve, CMS:2012qbp}. However, despite its success, some issues remain unexplained by the SM:  the lack of a dark matter candidate, the  hierarchy problem, unification, the flavor problem etc. This encourages the study of SM extensions. In the framework of the SM there are no tree-level flavor changing neutral currents (FCNC), which are, however, predicted by several SM extensions, being mediated by the Higgs boson or other new scalar or vector boson particles. In the context of these models, it is worth studying any signal that could give clues for new physics (NP), such as the widely studied process $\phi\to\tau\mu$, with $\phi$ a $CP$-even or $CP$-odd scalar boson \cite{Korner:1992zk,Diaz-Cruz:1999sns,Han:2000jz,Assamagan:2002kf,Arroyo-Urena:2013cyf,CMS:2021rsq,Huitu:2016pwk,Lami:2016mjf,Barradas-Guevara:2017ewn,Chamorro-Solano:2017toq,Primulando:2016eod,Arroyo-Urena:2018mvl,Arroyo-Urena:2020mgg,Hernandez-Tome:2020lmh,Urena:2021xtw}.  FCNC signals can also arise from the  top quark decays $t\to cX$ ($X=\phi, \gamma, g, Z, H$) \cite{Diaz-Cruz:1999wcs,Cordero-Cid:2004eih,Aguilar-Saavedra:2004mfd,Cordero-Cid:2005tjd,Kao:2011aa,Papaefstathiou:2017xuv,ATLAS:2018xxe,Arroyo-Urena:2019qhl,Gutierrez:2020eby}, and  from the less studied  decay of a new heavy scalar  boson into a top-charm quark pair \cite{Altunkaynak:2015twa}, which could be searched at the LHC and the  future high luminosity LHC  (HL-LHC). The latter aims to increase the LHC potential capacity  by reaching  a luminosity up to $\mathcal{L}=3000$ fb$^{-1}$ around 2035 \cite{Apollinari:2015wtw}. In this work we present a study of the $\phi\to tc$ decay in a  SM extension that incorporates a complex singlet $S_F$ via the Froggatt-Nielsen (FN) mechanism, which assumes that above some scale $\Lambda_F$ a symmetry (perhaps of Abelian type $U(1)_F$) forbids the Yukawa couplings with the SM fermions  charged under this symmetry; however, the Yukawa couplings can arise through non-renormalizable operators. The scalar spectrum of this model includes both a $CP$-even Flavon $H_F$ and a $CP$-odd Flavon $A_F$. The former can mix with the SM Higgs boson when the flavor scale is of the order of a few TeVs. A detailed study of the Flavon phenomenology can be consulted in Refs. \cite{Tsumura:2009yf,Berger:2014gga,Diaz-Cruz:2014pla,Bauer:2016rxs, Bolanos:2016aik}.
Our study not only  could serve as a strategy for the Flavon search, but it can also be  helpful  to assess the order of magnitude of flavor violation mediated by this particles, which is an indisputable signature of physics beyond the SM.

The organization of this paper is as follows: in Sec. \ref{sec2} we describe the most relevant theoretical aspects of the Froggatt-Nielsen singlet model (FNSM), which are necessary for our study. In Sec. \ref{sec3}  we  obtain the constraints on the  model parameters from  the most recent experimental results on the Higgs boson coupling modifiers $\kappa_i$ \cite{CMS:2018uag}, the full decay width of the Higgs boson \cite{Workman:2022ynf}, anomalous magnetic dipole moment of the muon \cite{Muong-2:2021ojo} and the perturbative limit. In addition, we include the current bound and the projections at the future colliders on ${\rm BR}(t\to ch)$ in order to constrain the $g_{\phi tc}$ coupling. Sec. \ref{sec4} is devoted to the study signal $pp\to \phi\to tc(t\to\ell\nu_{\ell}b)$ and the potential background as well as the strategy used to search for the  $\phi\to tc$ decay at the LHC and the HL-LHC. Finally, the conclusion are presented in Sec. \ref{sec5}.

\section{The Froggatt-Nielsen complex singlet model \label{sec2}}
We now focus on some relevant theoretical aspects of the FNSM. In Ref \cite{Bonilla:2014xba} a comprehensive analysis of the Higgs potential is presented, along with constraints on the parameter space from the constraints on the Higgs boson signal strengths  and the oblique parameters, including a few benchmark scenarios.  Also, the authors of   Ref. \cite{Barradas-Guevara:2017ewn} report a study  of the lepton flavor violating (LFV) Higgs boson decay $h\to\ell_i\ell_j$ in the scenario where there is $CP$ violation induced by a complex phase in the vacuum expectation value (VEV) of the complex singlet.
\subsection{The scalar sector}
In  addition to the SM-like Higgs doublet, $\Phi$, a FN complex singlet $S_F$ is introduced. They are given by
\begin{eqnarray}
& \Phi = \left( \begin{array}{c} G^+ \\ \frac{1}{\sqrt{2}} \left( v + \phi^0 + i G_z \right)\\
\end{array} \right), \qquad  \label{dec_doublets}&\\[2mm]
& S_F = \frac{1}{\sqrt{2}} (u + s + i p ), \label{dec_singlet}&
\end{eqnarray}
where $v$ is the SM VEV and $u$ is that of the FN complex singlet, whereas $G^+$ and $G^z$ are identified with the pseudo-Goldstone bosons that become the longitudinal modes of the $W^+$ and $Z$ gauge bosons.

We consider a scalar potential that respects a global $U(1)$ symmetry, with the Higgs doublet and the singlet transforming as $\Phi\to\Phi$ and $S_F\to e^{i\theta}S_F$. In general, such a scalar potential  admits a complex VEV, namely, $\langle S_F\rangle_0=ue^{-i\alpha}$, but in this work we consider the special case in which the Higgs potential is $CP$ conserving, i.e. we consider the limit with vanishing phase.
Such a $CP$-conserving Higgs potential is given by:
\begin{eqnarray}\label{potential}
V&=&-\frac{1}{2} m_1^2\Phi^{\dagger}\Phi-\frac{1}{2} m_{s_1}^2 S_F^*S_F - \frac{1}{2} m_{s_2}^2\left(S_F^{*2}+S_F^2\right) \nonumber\\
&+&\frac{1}{2}\lambda_1 \left(\Phi^{\dagger}\Phi\right)^2+\lambda_s\left(S_F^*S_F\right)^2
+\lambda_{11}\left(\Phi^{\dagger}\Phi\right)\left(S_F^* S_F\right),
\end{eqnarray}
where  $m_{s_2}^2$ stands for a $U(1)$-soft-breaking term, which is necessary to avoid the presence of a massless Goldstone boson, as will be evident below. Once the minimization conditions  are applied, the following relations are obtained:

\begin{eqnarray}
 m_{1}^2  &=&  v^2 \lambda_1 + u^2 \lambda_{11}, \\
 m_{s_1}^2 &=& -2 m^2_{s_2} + 2 u^2 \lambda_{s} + v^2 \lambda_{11}.
\end{eqnarray}

In this $CP$-conserving potential, the real and imaginary parts of the mass matrix do not mix. Thus, the mass matrix for the real components can be written in the  ($\phi^0,\,s$) basis as
\begin{equation}
M^2_S =
 \left( \begin{array}{cc}
   \lambda_1 v^2      &  \lambda_{11} u v \\
\lambda_{11} u v    &   2 \lambda_{s} u^2
\end{array} \right).
\end{equation}
The corresponding mass eigenstates are obtained via the standard $2\times 2$ rotation
\begin{eqnarray}\label{phi0_s}
 \phi^0   &=& \, \, \, \, \cos \, \alpha \, h + \sin \, \alpha  \, H_F, \nonumber \\
  s    &=& -\sin \, \alpha \, h + \cos \, \alpha \, H_F,
\end{eqnarray}
with $\alpha$ a mixing angle. Here $h$ is identified with the SM-like Higgs boson, with mass $m_h$=125 GeV, whereas the mass eigenstate $H_F$ is the $CP$-even Flavon.

As for the mass matrix of the imaginary parts, it is already diagonal in the  ($G_z , p$) basis:
\begin{equation}
M^2_P =
 \left( \begin{array}{cc}
   0       &  0     \\
   0      & 2 m^2_{s_2}
\end{array} \right),
\end{equation}
where the physical  mass eigenstate  $A_F=p$ is the $CP$-odd Flavon. Both $H_F$ and $A_F$ are considered to be  heavier than $h$.

\subsection{Yukawa sector}
The model, in addition to the new complex scalar singlet, also invokes the FN mechanism \cite{Froggatt:1978nt}.
The effective FN $U(1)_F$-invariant Lagrangian can be written as:
\begin{align}\label{Yuka}
 {\cal{L}}_Y &=    \rho^d_{ij} \left( \frac{ S_F }{\Lambda_F} \right)^{q_{ij}^d} \bar{Q}_{L_i}\Phi d_{R_j}  
                + \rho^u_{ij}  \left(\frac{ S_F }{\Lambda_F}\right)^{q_{ij}^u} \bar{Q}_{L_i}\tilde{\Phi} u_{R_j} 
                \nonumber\\&+ \rho^\ell_{ij}  \left(\frac{ S_F }{\Lambda_F}\right)^{q_{ij}^\ell} \bar{L}_{L_i} \Phi \ell_{R_j}   +{\rm H.c.},
\end{align}

which includes terms that  become the Yukawa couplings once the $U(1)$ flavor symmetry is spontaneously broken. Here $q_{ij}^f$ $(f=u,\,d,\,\ell)$ denote the charges of each fermion type under some unspecified Abelian flavor symmetry, which help to explain the fermion mass hierarchy; $\rho^f_{ij}$ are dimensionless couplings seemingly of $\mathcal{O}(1)$, $\Lambda_F$ represents the flavor scale and
\begin{eqnarray}\label{Scalar-fields}
\bar{Q}_{L_i}^T&=&(u_{L_i},\,d_{L_i}),\nonumber\\
\bar{L}_{L_i}^T&=&(\nu_{L_i},\,\ell_{L_i}),\\
\tilde\Phi&=&i\sigma^2\Phi^*.\nonumber
\end{eqnarray}
  
We now write the neutral component of the Higgs field in the unitary gauge  and  use the first order expansion
\begin{equation}
\Bigg(\frac{S_F}{\Lambda_F}\Bigg)^{q_{ij}}=\left(\frac{u+s+ip}{\sqrt{2}\Lambda_F}\right)^{q_{ij}}
\simeq\left(\frac{u}{\sqrt{2}\Lambda_F}\right)^{q_{ij}}\left[1+q_{ij}\left(\frac{s+ip}{u}\right)\right],
\end{equation}
along with Eqs. \eqref{dec_singlet}, \eqref{phi0_s} and \eqref{Scalar-fields}. We also define $Y_{ij}^f=\rho_{ij}^f(u/\sqrt{2}\Lambda_F)^{q_{ij}^f}$,  $\tilde{M}^f=(v/\sqrt{2})Y_{ij}^f$, $r_s=v/(\sqrt{2}u)$. In order to diagonalize the mass matrix $\tilde{M}^f$, the electroweak fields are redefined as
\begin{eqnarray}
F_L\to U_L^f F_L,\,f_R\to U_R^f f_R\Rightarrow Y^f=U_L^{f\dagger}Y^f_{\text{diago}}U_R^f,
\end{eqnarray}
where $Y^{f=\ell}_{\text{diago}}=\frac{\sqrt{2}}{v}\text{diago}(m_e,\,m_{\mu},\,m_{\tau})=\frac{\sqrt{2}}{v}M^{\ell}$, analogously for the case of quarks.  Thus,  one gets the following Yukawa Lagrangian for the Higgs- and Flavon-fermion interactions:
\begin{eqnarray}\label{Yukalagrangian}
{\cal{L}}_Y &=& \frac{1}{v}[\bar{U} M^u U+\bar{D} M^d D+\bar{L}M^{\ell}L](c_{\alpha}h+s_{\alpha}H_F) \nonumber\\
&+&r_s[\bar{U}_i\tilde{Z}^u U_j+\bar{D}_i\tilde{Z}^d D_j+\bar{L}_i\tilde{Z}^{\ell} L_j]
(-s_{\alpha}h+c_{\alpha}H_F+iA_F)+{\rm H.c.},
\end{eqnarray}
where $s_{\alpha}\equiv \sin\alpha $, $c_\alpha\equiv \cos\alpha$. A fact to highlight is that the intensity of the flavor violating (FV) couplings are encapsulated in the $\tilde{Z}_{ij}^f=U_L^{{f\dagger}}Z_{ij}^fU_R^f$ matrices. In the flavor basis, the $Z_{ij}^f$ matrix elements are given by
$Z_{ij}^f=\rho_{ij}^f (u/\sqrt{2}\Lambda_F)^{q_{ij}^f}q_{ij}^f$, which remains non-diagonal even after diagonalizing the mass matrices, thereby giving rise to FV scalar couplings. In addition to the Yukawa couplings, we  also need  the $\phi VV$ ($V=W,\,Z$)  couplings for our calculation, which can be extracted from the kinetic terms of the Higgs doublet and the complex singlet.
In Table \ref{couplings} we show the coupling constants for the interactions of  the SM-like Higgs boson and the Fla\-vons to  fermions and gauge bosons.
\begin{table}
\caption{Couplings of the SM-like Higgs boson $h$ and the Flavons $H_F$ and $A_F$ to fermion pairs and gauge boson pairs in the FNSM. Here $r_s=v/\sqrt{2}u$.\label{couplings}}
\begin{centering}
\begin{tabular}{cc}
\hline
\hline
Vertex ($\phi XX$) &Coupling constant ($g_{\phi XX}$) \tabularnewline
\hline
\hline
${hf_{i}\bar{f}_{j}}$ & $\frac{c_{\alpha}}{v}M_{ij}^{f}-s_{\alpha}r_{s}\tilde{Z}^{f}_{ij}$\tabularnewline
${H_{F}f_{i}\bar{f}_{j}}$ & $\frac{s_{\alpha}}{v}M_{ij}^{f}+c_{\alpha}r_{s}\tilde{Z}^{f}_{ij}$\tabularnewline
${A_{F}f_{i}\bar{f}_{j}}$ & $ir_{s}\tilde{Z}^{f}_{ij}$\tabularnewline
${hZZ}$ & $\frac{gm_{Z}}{c_{W}}c_\alpha$\tabularnewline
${hWW}$ & $gm_{W}c_\alpha$\tabularnewline
${H_{F}ZZ}$ & $\frac{gm_{Z}}{c_{W}}s_\alpha$\tabularnewline
${H_{F}WW}$ & $gm_{W}s_\alpha$\tabularnewline
\hline
\hline
\end{tabular}
\par\end{centering}
\end{table}

\section{Constraints on the FNSM parameter space \label{sec3}}
To evaluate the decay widths and production cross-sections of the Flavons $H_F$ and $A_F$, we need the bounds on the parameter space of our model, they are: 
\begin{itemize}
\item The mixing angle $\alpha$.
\item The VEV of the FN complex singlet $u$.
\item The matrix element $\textbf{$\tilde{Z}_{tc}$}$.
\item The Flavon masses $m_{H_F}$ and $m_{A_F}$.
\end{itemize}
\subsection{Constraint on the mixing angle $\alpha$ and VEV of singlet $u$}
It turns out that these parameters can be constrained via  the Higgs boson coupling modifiers  $\kappa_j$ ($j= W,\,Z,\,g,\,b,\,\tau,\,\mu$) \cite{CMS:2018uag}, which are defined for a given Higgs boson production mode $i\to h$ or decay channel $h\to j$ as
\begin{equation}
\kappa_i^2=\sigma_i/\sigma_i^{\rm SM}\qquad \text{or}\qquad \kappa_j^2=\Gamma_j/\Gamma_j^{\rm SM},
\end{equation}
{where $\sigma_i^{\rm SM}$ ($\Gamma_j^{\rm SM}$) stands for the pure SM contributions, whereas $\sigma_i$ ($\Gamma_j$)  includes new physics contributions. 

Figure \ref{calpha-u}(a) shows the $c_{\alpha}-u$ plane, where each colored area represents the allowed regions by $\kappa_j$ considering the expected results at the HL-LHC at a confidence level of $2\sigma$. Besides, in the same plot, the intersection of all $\kappa_j's$ is included, which coincides with $\kappa_{\tau}$ since the latter is the most restrictive. Meanwhile, we present separately in Fig. \ref{calpha-u}(b) the intersection of all $\kappa_j's$ and the allowed region by both the perturbative limit applied on the parameter of the potential $\lambda_s=(m_{A_F}^2 + c_{\alpha}^2 m_{H_F}^2 + m_h^2 s_{\alpha}^2)/(2 u^2)\leq 4\pi$ and the current discrepancy between the experimental measurement and the SM theoretical prediction \cite{Muong-2:2021ojo} of the anomalous magnetic dipole moment given by
\begin{eqnarray}
\Delta a_{\mu}&=&(25.1\pm5.9)\times 10^{-10},\\ \nonumber
\Delta a_{\mu}^{\text{FNSM}}&\approx&\frac{m_{\mu}}{16\pi^2}\sum_{\phi=h,\,H_F,\,A_F}\sum_{\ell=\mu,\,\tau}\frac{m_{\ell}g_{\phi\mu\ell}^2}{m_{\phi}}\Bigg(2\ln\Bigg(\frac{m_{\phi}^2}{m_{\ell}^2}\Bigg)-3\Bigg).
\end{eqnarray}

We notice in Fig. \ref{calpha-u}(b) that $c_{\alpha}$ is close to unity, this is to be expected because the dominant term of the $g_{hf_i\bar{f}_i}$  coupling in Table \ref{couplings} is proportional to $c_{\alpha}$. When $c_{\alpha}=1$, the SM case is recovered. As far as the VEV of the FN complex singlet is concerned, it is a lower limit imposed by the perturbative limit; the most stringent is when $m_{A_F}=m_{H_F}=1000$ GeV, $u\geq 281$ GeV. The exploration of the muon anomalous magnetic dipole moment help us to find a upper limit on $u\leq 1100$ GeV, in addition to imposing a lower limit on $c_{\alpha}\geq 0.995$. We also explored the total decay width of the Higgs boson in order to find additional constrains on the mixing angle $\alpha$ and $u$, however this observable is not restrictive.

\begin{figure}[h!]
\centering
    {\subfigure[]{\includegraphics[scale = 0.25]{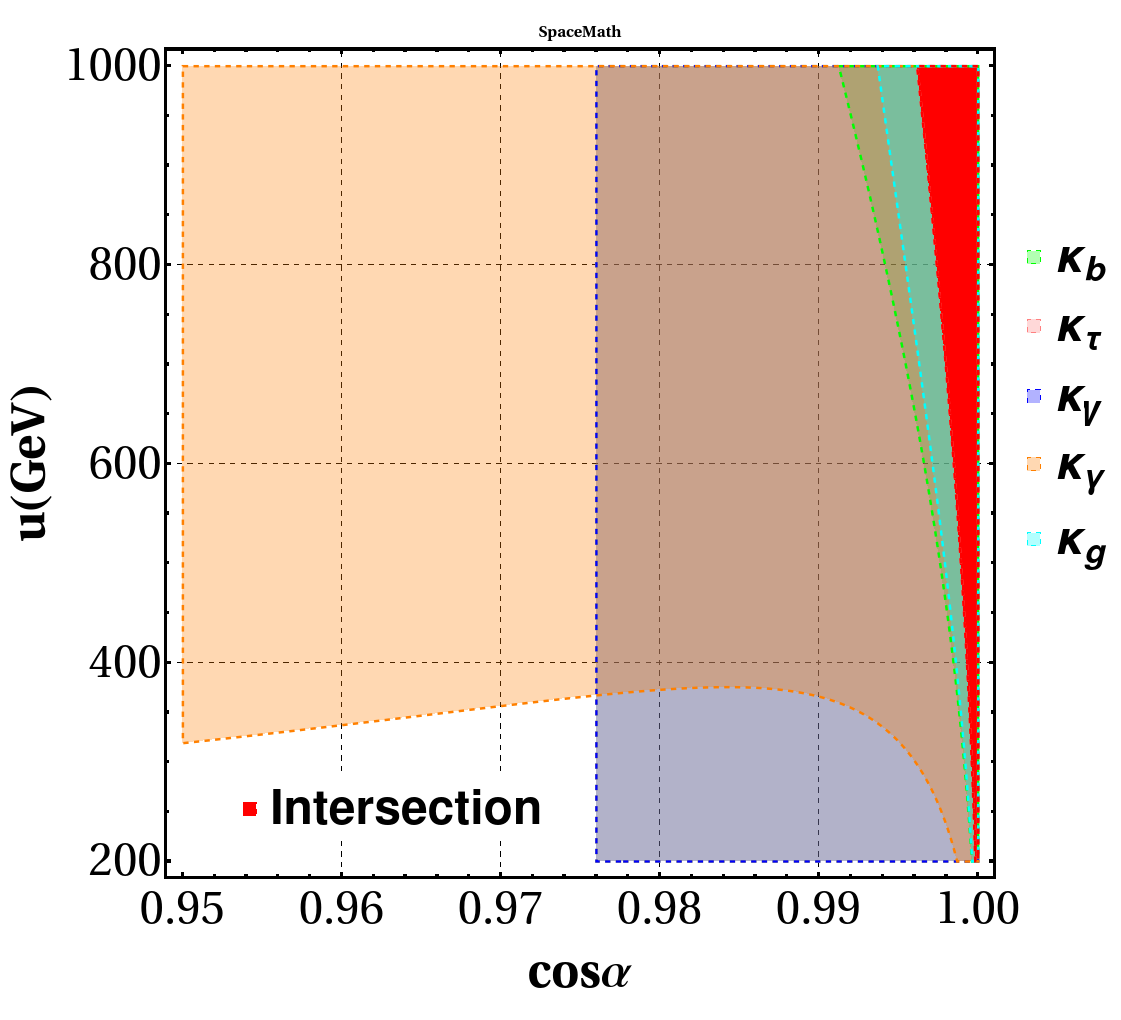}}}
    {\subfigure[]{\includegraphics[scale = 0.19]{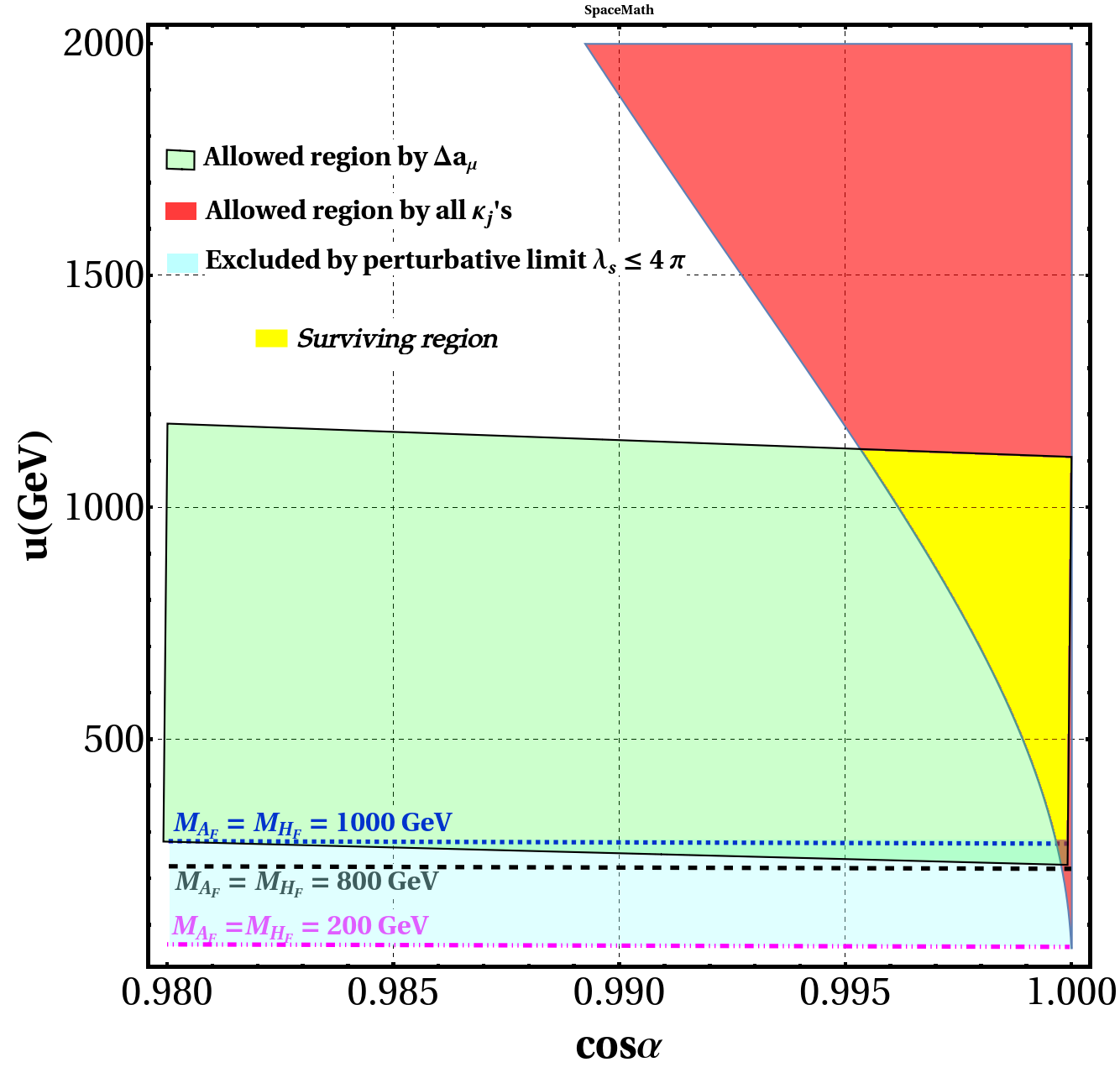}}}
    \caption{{(a) Allowed regions by all $\kappa_j$ coupling modifiers in the $c_\alpha-u$ plane, where $V=Z,\,W$; (b) Only intersection of $\kappa_j's$ and excluded zone by perturbative limit.
        } \label{calpha-u} }
	\end{figure}

\subsection{Constraint on $\tilde{Z}_{tc}$}	
	
So far, we only have considered the  bound on the diagonal couplings; however, we need a bound on the $\tilde{Z}_{tc}$ matrix element in order to evaluate  the $\phi\to tc$ decay.  To our knowledge, there are no processes from which we can extract a stringent bound on  $\tilde{Z}_{tc}$, but we can assess its order of magnitude by considering { the upper limits ${\rm BR}(t\to ch)<1.1\times 10^{-3}$ \cite{Workman:2022ynf}. We also consider the  prospect for the branching ratio ${\rm BR}(t\to ch)<4.3\times 10^{-5}$ searches at the FCC-hh \cite{Mandrik:2018yhe}. This is show in   Fig. \ref{c}.
\begin{figure}[!htb]
\centering
 \subfigure{\includegraphics[scale = 0.5]{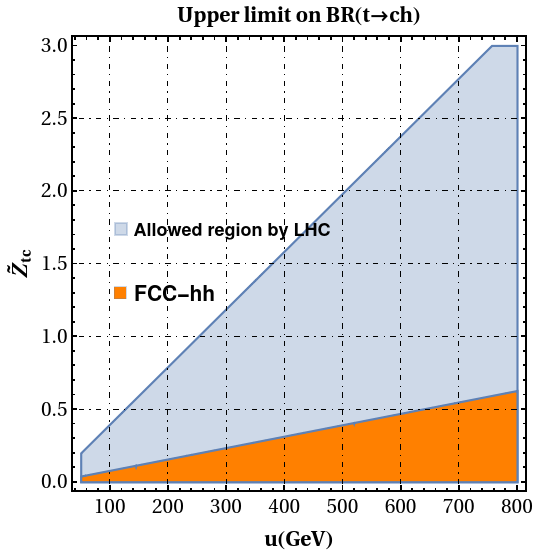}}
 \caption{Allowed region in the $u$-$\tilde Z_{tc}$ plane from the current bound on ${\rm BR}(t\to ch)<1.1\times 10^{-3}$ (blue color) and the projection at the FCC-hh (orange color).\label{c}}
	\end{figure}
	
As for the bounds on the $\tilde{Z}^{\ell\ell}$ diagonal matrix elements, we use those obtained in Ref. \cite{Arroyo-Urena:2018mvl}. We summarize in  Table \ref{TablaParametros}  the values of the FNSM parameters used in the evaluations; while in Table \ref{TablaBMP} we define three benckmark points to be used in the Monte Carlo simulation.
\begin{table}
\caption{Model parameter values considered in the numerical analysis. \label{TablaParametros}}
\begin{centering}
\begin{tabular}{cc}
\hline
\hline
Parameter & Value\tabularnewline
\hline
\hline
$c_{\alpha}$ & $0.999$\tabularnewline
$u$   & $600$ and $1000$ (GeV)\tabularnewline
$\tilde{Z}_{tt}$ & $0.5$\tabularnewline
$\tilde{Z}_{bb}$ & $0.1$\tabularnewline
$\tilde{Z}_{tc}$ & $0.05$, $0.2$ and $0.45$\tabularnewline
$\tilde{Z}_{\tau\tau}$ & $0.1$ \cite{Arroyo-Urena:2018mvl}\tabularnewline
$\tilde{Z}_{\mu\mu}$ & $10^{-3}$ \cite{Arroyo-Urena:2018mvl}\tabularnewline
$\tilde{Z}_{\tau\mu}$ & $0.35$ \tabularnewline
$m_{A_{F}}$ & $0.2-1$ (TeV) \tabularnewline
$m_{H_{F}}$ & $0.2-1$ (TeV)\tabularnewline
\hline
\hline
\end{tabular}
\par\end{centering}
\end{table}

\begin{table}
\caption{Benckmark points used in the  Monte Carlo simulation.\label{TablaBMP}}
\begin{centering}
\begin{tabular}{cc}
\hline
\hline
Benchmark points (BMP) \tabularnewline
\hline
\hline
BMP1: $\tilde{Z}_{tc}=0.45,\,u=600,\,1000$ GeV\tabularnewline
BMP2: $\tilde{Z}_{tc}=0.2,\,u=600,\,1000$ GeV\tabularnewline
BMP3: $\tilde{Z}_{tc}=0.05,\,u=600,\,1000$ GeV\tabularnewline
\hline
\hline
\end{tabular}
\par\end{centering}
\end{table}


\section{Search for $\phi\to tc$ decays at the HL-LHC \label{sec4}}

\subsection{Flavon decays}
We now present the behavior of the branching ratios of the main Flavon decay channels, which were obtained via our own \texttt{Mathematica} package so-called \texttt{SpaceMath} \cite{Arroyo-Urena:2020qup}, that implements the analytical expressions for the corresponding decay widths. A cross-check was done by comparing  our results with those obtained via \texttt{CalcHEP} \cite{Belyaev:2012qa}, in which we implemented the corresponding Feynman rules via the \texttt{LanHEP} package \cite{Semenov:2014rea}. In  Fig. \ref{BAXX} we show the branching ratios  of the $CP$-odd   Flavon $A_F$ as  functions of its mass $m_{A_F}$; we use the parameter values of Table \ref{TablaParametros}. As $A_F$ does not couple to gauge bosons at tree-level, its dominant decay modes are  $A_F\to t\bar{t}$, $A_F\to \tau^-\mu^+$,  and $A_F\to t\bar{c}$, with a branching ratio at the $\mathcal{O}(0.1)$ level for masses of the Flavon $A_F$ in the $200\leq m_{A_F}\leq 1000$ GeV. Other interesting channels such as $A_F\to gg$ and $A_F\to b\bar{b}$ search a branching ratio of $\mathcal{O}(10^{-3})-\mathcal{O}(10^{-2})$.

 \begin{figure}[!htb]
\centering
    \subfigure[ ]{\includegraphics[width=8cm]{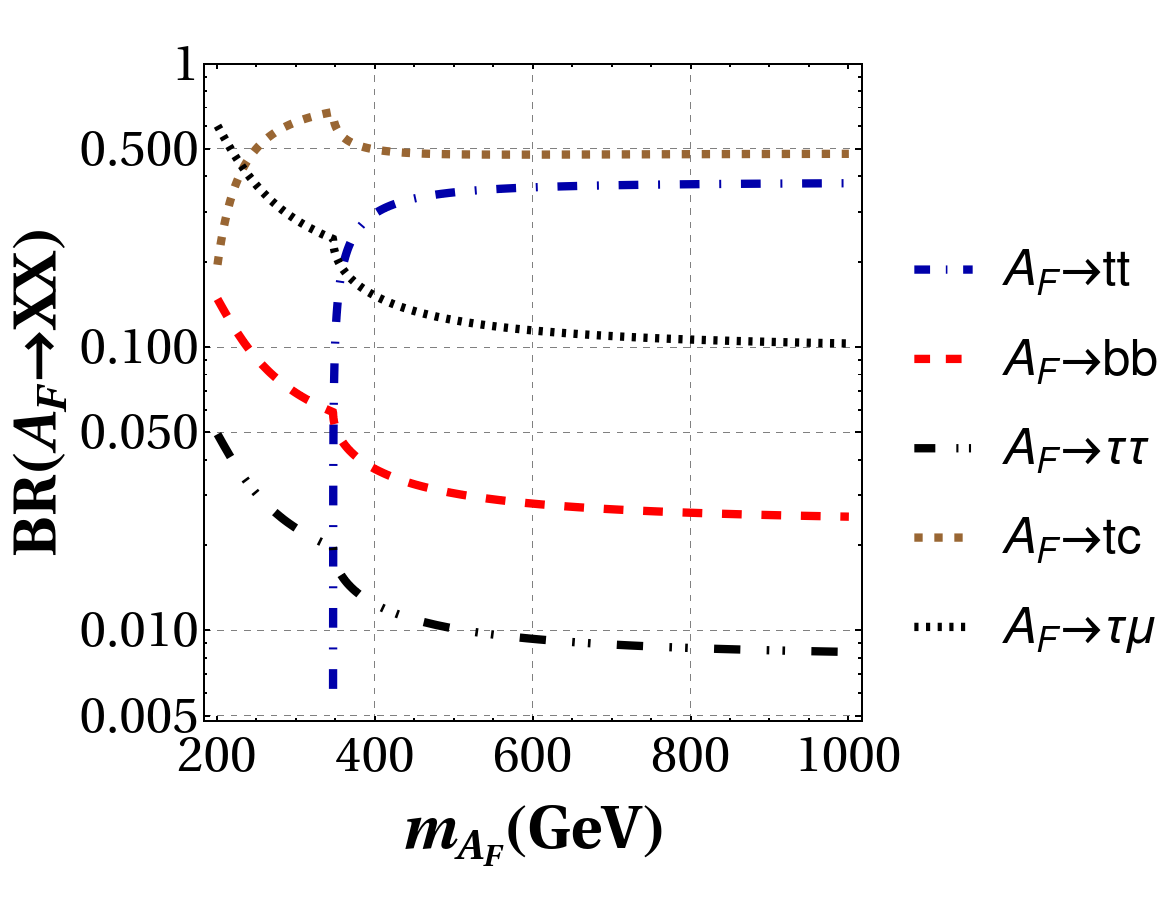}}
 \subfigure[ ] {\includegraphics[width=8cm]{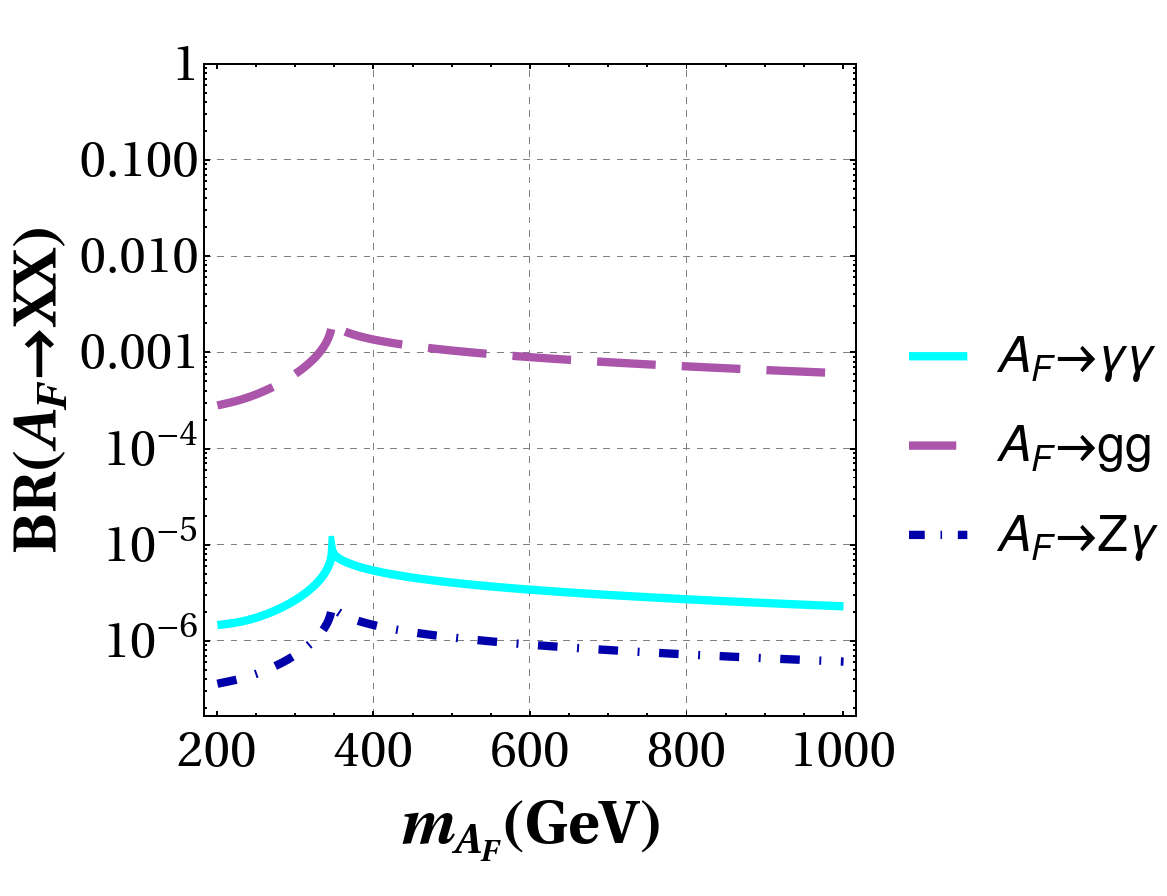}}
     \caption{Branching ratios of the two-body decay modes of a $CP$-odd flavon as a function of its mass for { the parameter values of Table \ref{TablaParametros} ($u=1000$ GeV and $\tilde{Z}_{tc}=0.45$).}\label{BAXX}}
 	\end{figure}

 As far as the $CP$-even Flavon $H_F$ is concerned, the branching ratios for their main decay channels are presented in Fig. \ref{BHXXFV}, for the same parameter values used for the $A_F$ decays. We observe that the dominant $H_F$ decay channels are $H_F\to \tau^-\mu^+$ and $H_F\to t\bar{c}$ for $m_{H_F}\leq 2m_{top}$, with branching ratios of order $\mathcal{O}(10^{-1})$. Another important channel is $H_F\to hh(h\to\gamma\gamma,\,h\to b\bar{b})$ which was studied by one of the authors of this project in Ref. \cite{Arroyo-Urena:2022oft}.
Conversely, when $m_{H_F}\geq 2m_{top} $, the dominant channels are $H_F\to t\bar{t},\, W^+W^-,\,ZZ$ and $\tau^-\mu^+$. Other decay modes such as $H_F\to b\bar{b}$, $H_F\to \tau^-\tau^+$, $H_F\to \gamma\gamma$ and $H_F\to gg$ have branching ratios ranging from $10^{-6}$ to $10^{-3}$, whereas the decays $H_F\to Z\gamma$ and $H_F\to \mu\mu$ are very suppressed. 

 \begin{figure}[!htb]	
 \centering
 \subfigure[ ]{\includegraphics[width=8.3cm]{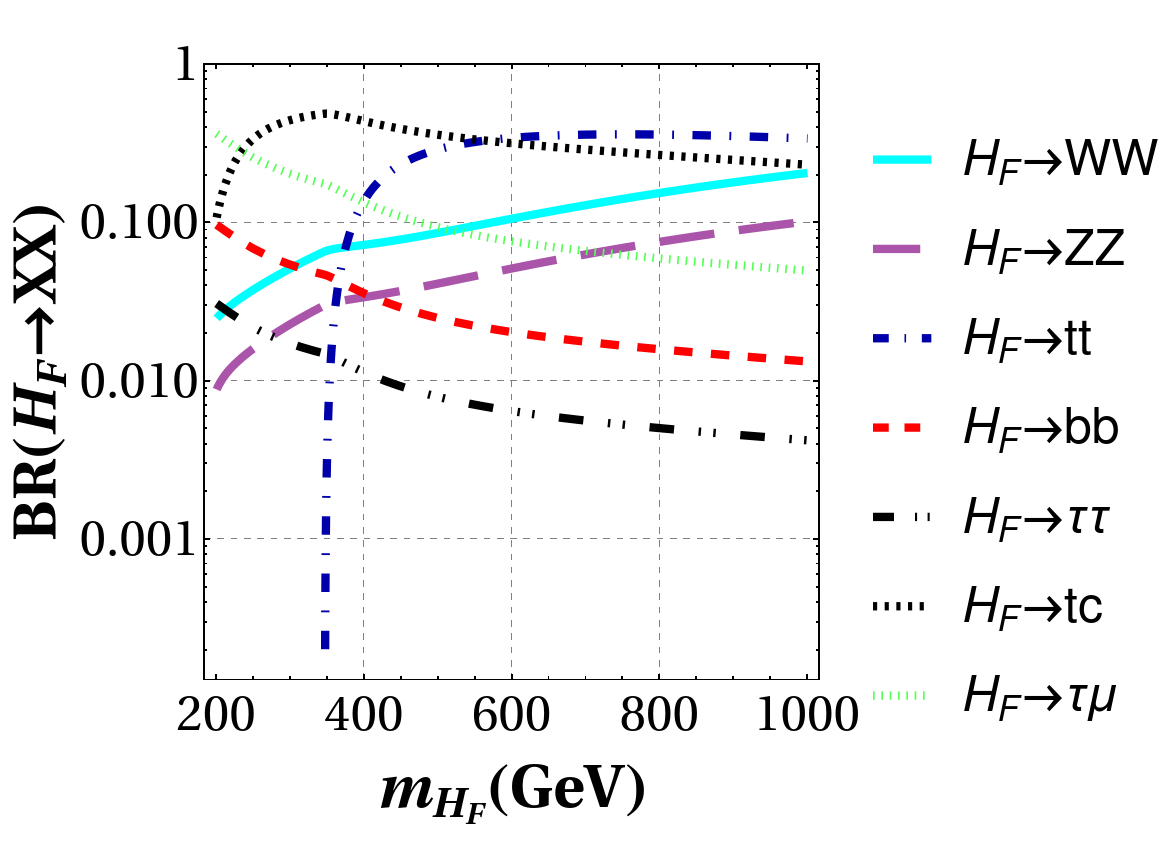}}
 \subfigure[ ] {\includegraphics[width=7.7cm]{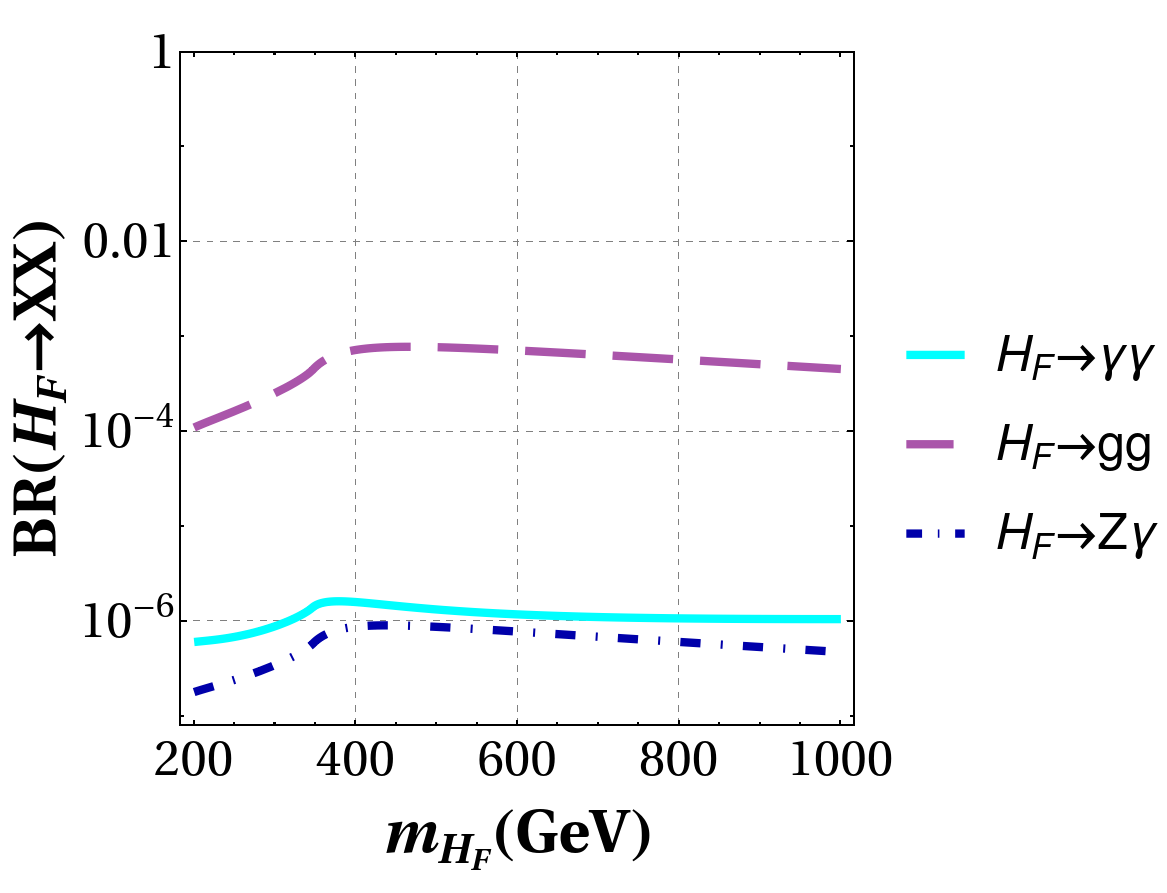}}
 \caption{Branching ratios of the two-body decay modes of a $CP$-even flavon as a function of its mass for the parameter values of Table \ref{TablaParametros} ($u=1000$ GeV and $\tilde{Z}_{tc}=0.45$).\label{BHXXFV}}
\end{figure}

\subsection{Events}
In this section we now present a Monte Carlo analysis for  the production of both the $H_F$ and the $A_F$ Flavons at the LHC via gluon fusion  $gg\to \phi$ ($\phi=H_F,\, A_F$), followed by the  FCNC decay  $\phi \to tc$. We  apply realistic  kinematic cuts  and  consider tagging and miss tagging efficiencies. We then obtain the statistical significance, which  could be experimentally confirmed.

We present in  Fig. \ref{Eventos} the number of events produced $\sigma(gg\to \phi\to tc\, (t\to \ell\nu_{\ell}b))\times \mathcal{L}$ $(\equiv\mathcal{N}_{\phi})$, where $\mathcal{L}=300$ fb$^{-1}$ is the integrated luminosity at the final stage of the LHC. For this computation, we use \texttt{CalcHEP} \cite{Belyaev:2012qa} with the CT10 parton distribution functions \cite{Gao:2013xoa}. We note that for both Flavon masses $m_{\phi}$, $\mathcal{N}_{\phi}$ is similar in the $400\leq m_{\phi}\leq 1000$ GeV interval. Meanwhile, for masses in the range $200\leq m_{\phi}\leq 350$ GeV, $\mathcal{N}_{A_F}\approx 3\mathcal{N}_{H_F}$. These results are encouraging since similar statistical significance will be obtained, despite different kinematic behaviors.

\begin{figure}[!htb]
\centering
\includegraphics[scale = 0.26]{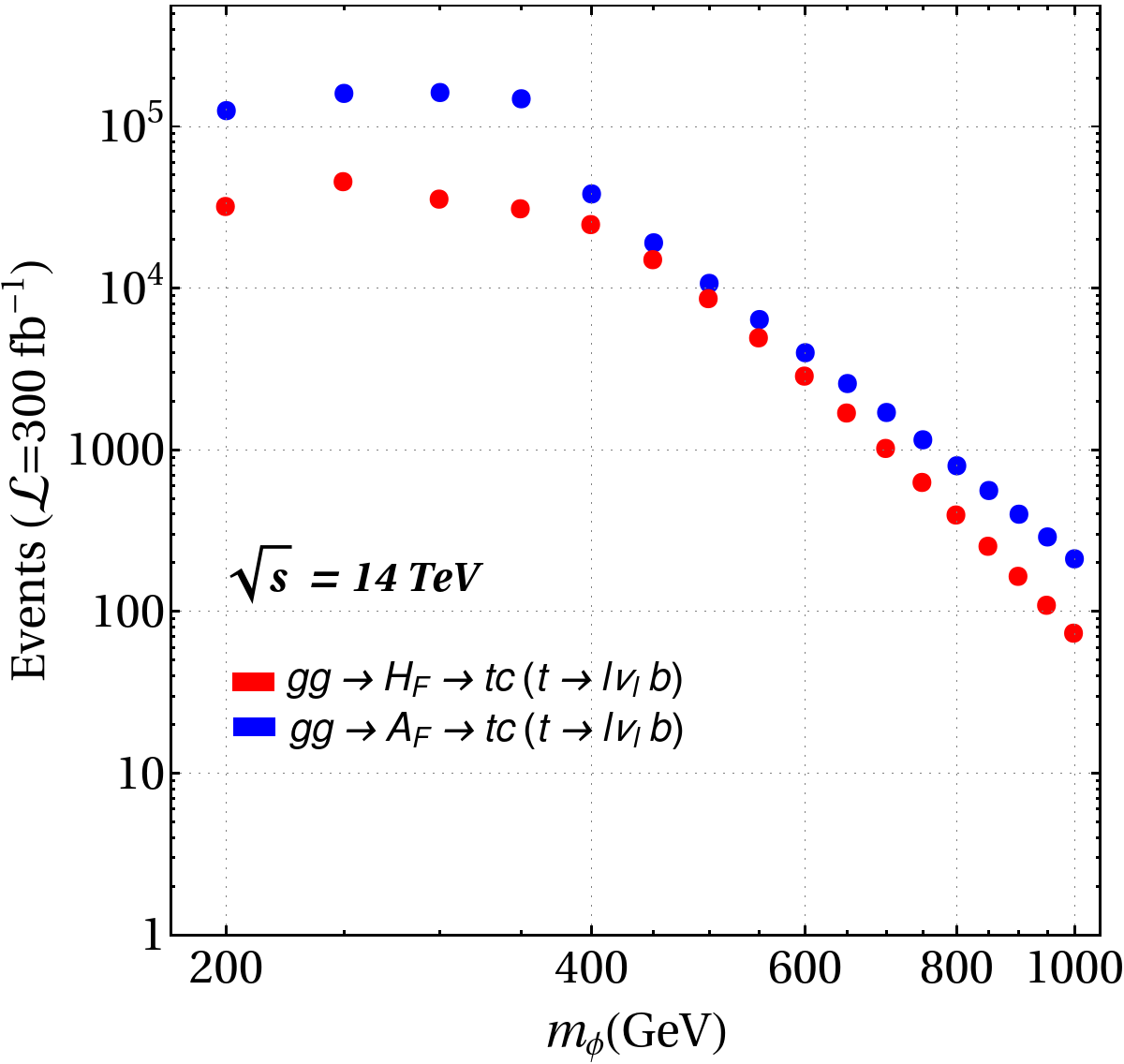}
    \caption{Number of events produced for the process $gg\to\phi \to tc\,(t\to \ell\nu_{\ell}b)$ as a function of the  Flavon mass $m_{\phi}$ at $\sqrt{s}$=14 TeV with an integrated luminosity of $\mathcal{L}=300$ fb$^{-1}$. }  \label{Eventos}
	\end{figure}

\subsubsection{Kinematic cuts}

We now turn to the Monte Carlo simulation, for which we use \texttt{Madgraph5} \cite{Alwall:2011uj}, with the corresponding Feynman rules generated via \texttt{LanHEP} \cite{Semenov:2014rea} for a \texttt{UFO} model \cite{Degrande:2011ua}. To perform shower and hadronization we use \texttt{Pythia8} \cite{Sjostrand:2014zea}. 

The signal and the main background events are as follows:
\begin{itemize}
\item \textbf{SIGNAL:} The signal is $gg\to \phi \to tc\to b\ell\nu_{\ell}c$ with $\ell=e,\,\mu$. We generated $10^5$ events scanning over $m_{\phi}\in [200,\,1000]$ TeV and considered  the parameter values of  Table \ref{TablaParametros}.

\item \textbf{BACKGROUND:} The dominant SM background arises from the final states $Wjj+Wb\bar{b}$, $tb+tj$ and $t\bar{t}$, in which either one of the two leptons is missed in the semi-leptonic top quark decays or two of the four jets are missed when one of the top quarks decays semi-leptonically. 
\end{itemize}

In Fig. \ref{distributions} we present the kinematic distributions generated both by the background processes and the decay of $A_F$ for $m_{A_F}=200$ GeV, namely, the transverse momentum of the particles produced by the decay of the top quark: (a) leading b-jet, (b) the charged lepton, (c) the missing energy transverse (MET) due to the neutrino in the final state is displayed. The transverse momentum of the leading jet is shown in (d). Finally, the transverse masses of the top quarks and CP-odd Flavon are depicted in (e) and (f). Meanwhile, in Figs. \ref{distributionsSGN1}, \ref{distributionsSGN2}, \ref{distributionsSGN3} shows the same as in Fig. \ref{distributions} but only for the signal to $m_{A_F}=200,\,400,\,900$ GeV.  
\begin{figure*}[!htb]	
 \centering
 \subfigure[ ]{\includegraphics[width=5.5cm]{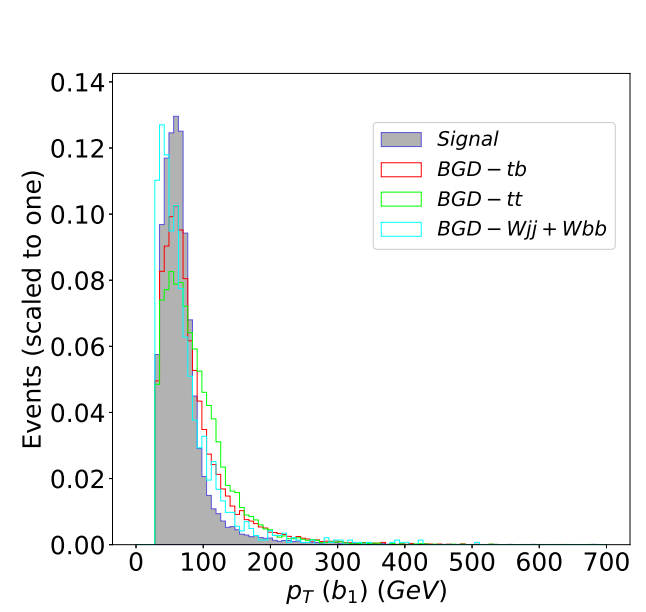}}
 \subfigure[ ] {\includegraphics[width=5.5cm]{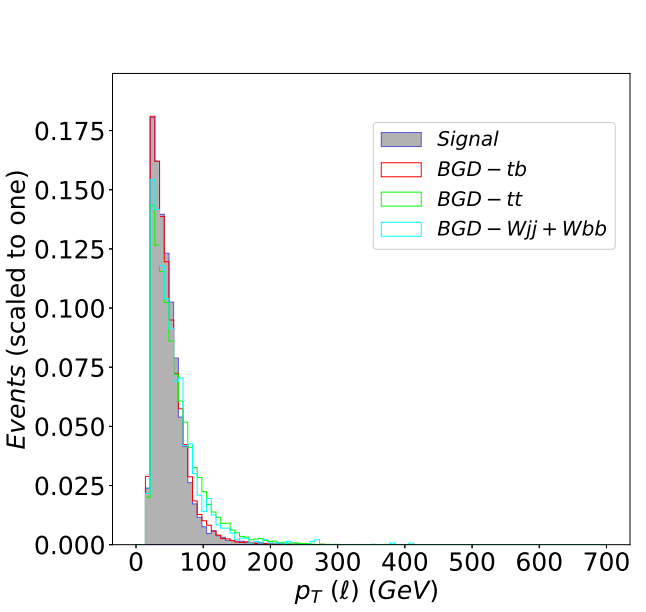}}
 \subfigure[ ] {\includegraphics[width=5.5cm]{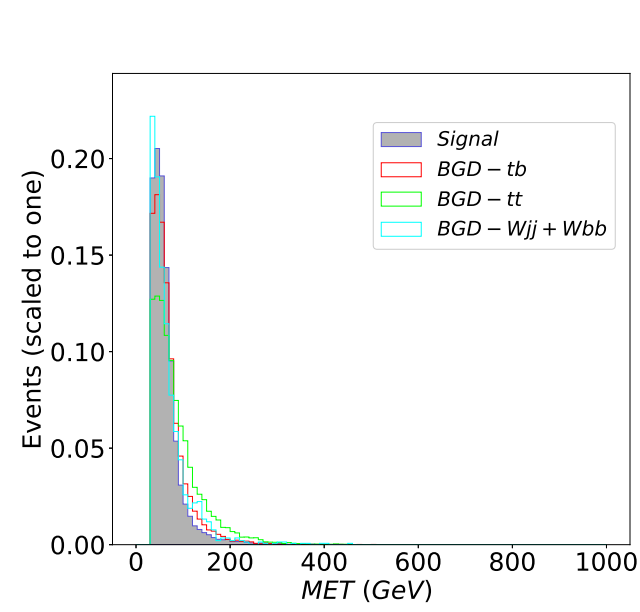}}
 \subfigure[ ] {\includegraphics[width=5.5cm]{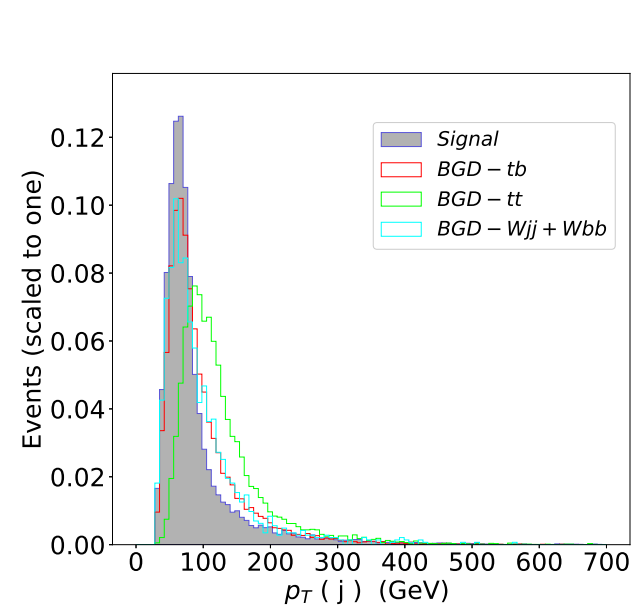}}
 \subfigure[ ] {\includegraphics[width=5.5cm]{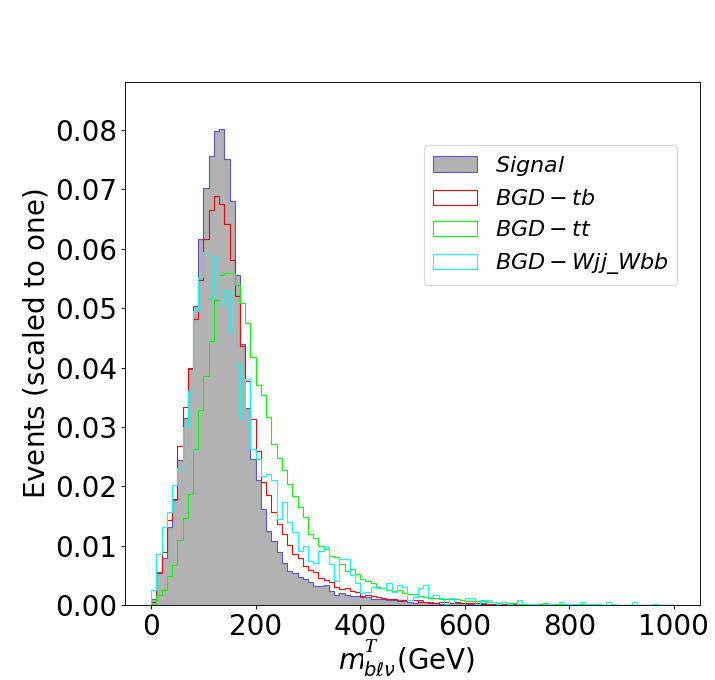}}
 \subfigure[ ] {\includegraphics[width=5.5cm]{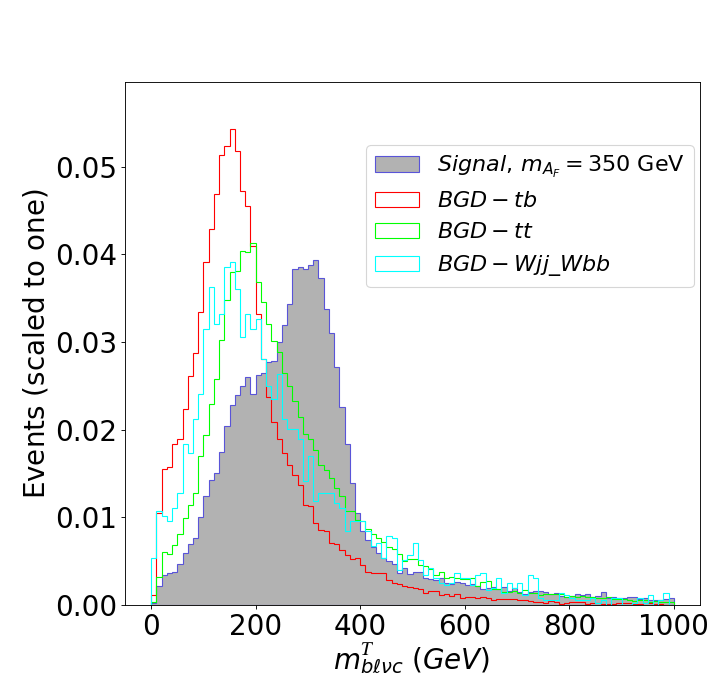}}
 \caption{Normalized transverse momentum distributions associated to the top decay: (a) leading b-jet, (b) leading charged lepton; (c) tranverse missing energy due to undetected neutrinos; (d) transverse momentum distribution of the c-jet; (e) top quark transverse mass $(m_{b\ell \nu }^T)$ and (f) CP-odd Flavon transverse mass $(m_{b\ell \nu c}^T)$ considering $m_{A_F}=350$ GeV.}\label{distributions}
\end{figure*}

The  kinematic cuts imposed to study a possible evidence of the $\phi\to tc$ ($m_{\phi}=200$ GeV) at the LHC are as follows.
\begin{enumerate}
\item  We requiere two jets with $|\eta^j|<2.5$ and $p_T^j>30$ GeV, one of them is tagged as a $b$-jet.

\item  We require one isolated lepton ($e\,\text{or}\,\mu$) with $|\eta^{\ell}|<2.5$ and $p_T^{\ell}>20$ GeV.

\item Since an undetected neutrino is included in the final state, we impose the cut  MET$>30$ GeV.

\item Finally, we impose a cut on the transverse masses $m_{b\ell \nu c}^T$ and $m_{b\ell \nu }^T$ as follows
\begin{itemize}
\item $0.8 m_{A_F}<m_{b\ell \nu c}^T<1.2m_{A_F}$,
\item $0.8 m_{\text{top}}<m_{b\ell \nu }^T<1.2m_{\text{top}}$. 
\end{itemize}
\end{enumerate}

 The kinematic analysis was done via \texttt{MadAnalysis5} \cite{Conte:2012fm} and for detector simulations we use \texttt{Delphes} \cite{deFavereau:2013fsa}. As far as the jet reconstruction, we use the jet finding package \texttt{FastJet} \cite{Cacciari:2011ma} and the anti-$k_T$ algorithm \cite{Cacciari:2008gp}. We include also the tagging and misstagging efficiencies $b$-tagging efficiency $\epsilon_b=90\%$ and to account for the probability that a $c$-jet is miss tagged as a $b$-jet we consider $\epsilon_c=10\%$, whereas for any other jet we use $\epsilon_j=1\%$.
\begin{figure}[!htb]	
 \centering
 \subfigure[ ]{\includegraphics[width=8cm]{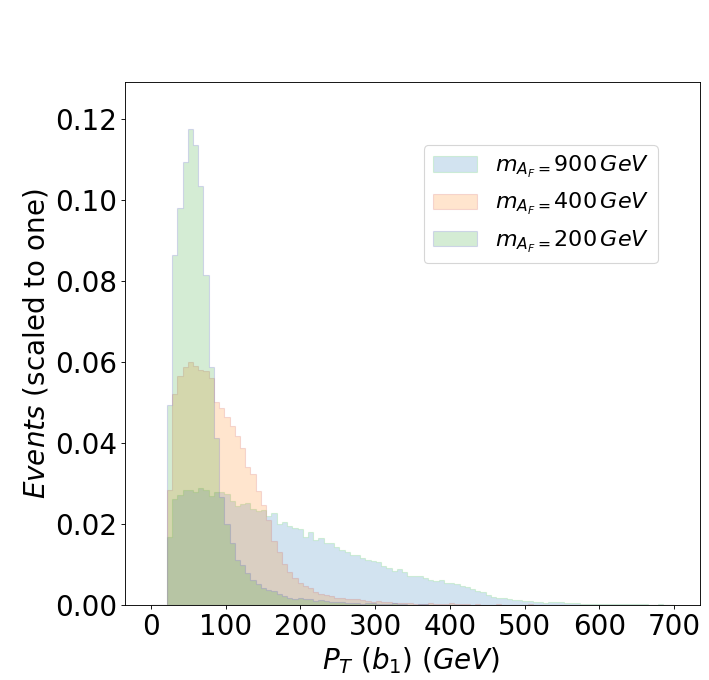}}
  \subfigure[ ] {\includegraphics[width=8cm]{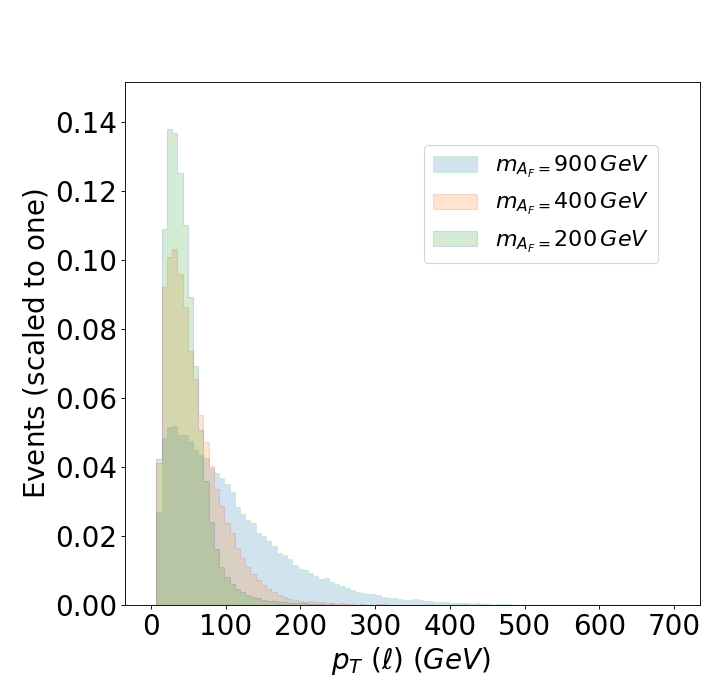}}
   \caption{Normalized distributions generated by the decay of $A_F$ for $m_{A_F}=200,\,400,\,900$ GeV. Transverse momentum of (a) leading b-jet and (b) leading charged lepton.}\label{distributionsSGN1}
\end{figure}

\begin{figure}[!htb]	
 \centering
   \subfigure[ ] {\includegraphics[width=8cm]{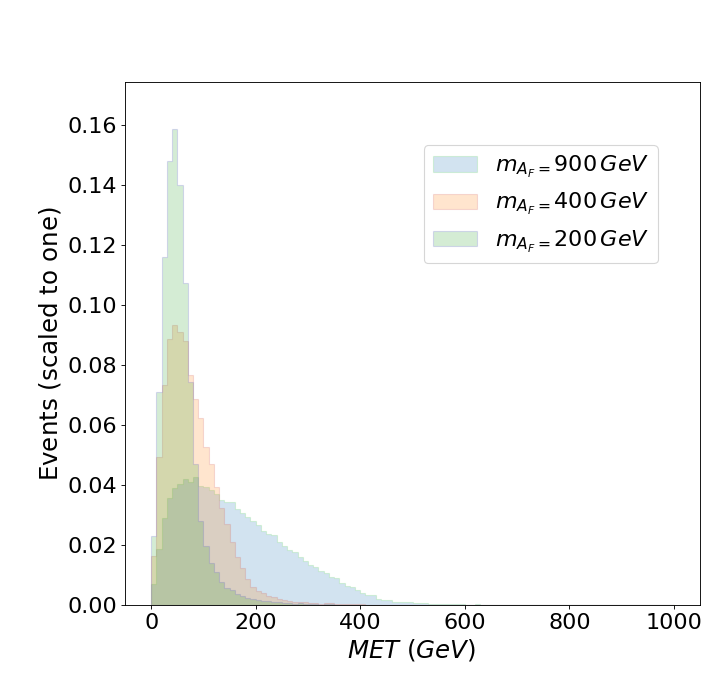}}
 \subfigure[ ] {\includegraphics[width=8cm]{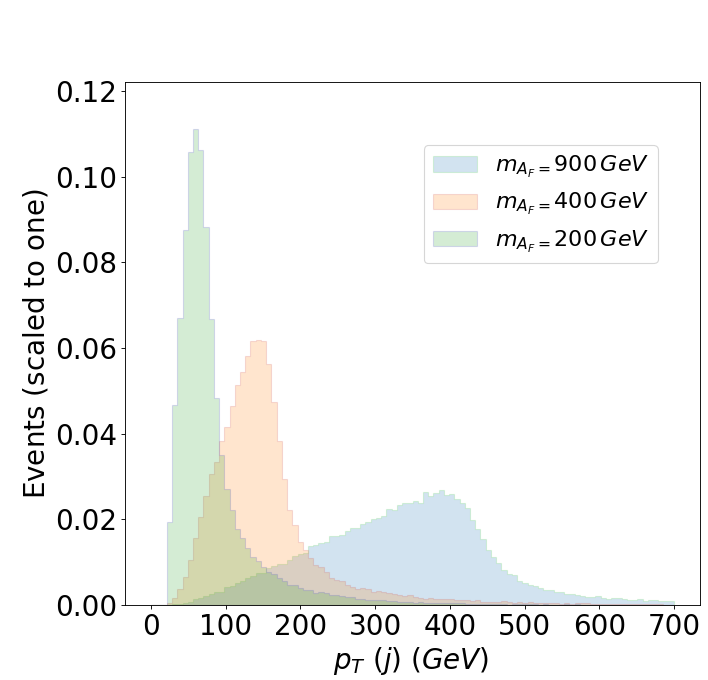}}
 \caption{Normalized distributions generated by the decay of $A_F$ for $m_{A_F}=200,\,400,\,900$ GeV. (a) transverse missing energy due to undetected neutrino, (b) transverse momentum distribution of the c-jet.}\label{distributionsSGN2}
\end{figure}
\begin{figure}[!htb]	
 \centering
 \includegraphics[width=8cm]{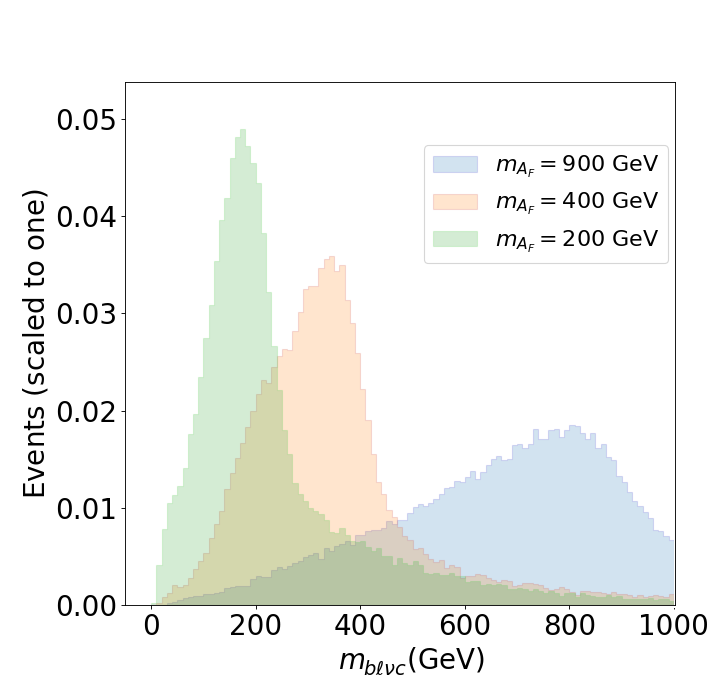}
 \caption{Reconstructed CP-odd Flavon mass for $m_{A_F}=200,\,400,\,900$ GeV.}\label{distributionsSGN3}
\end{figure}

We now compute the signal significance $\mathcal{S}=N_S/\sqrt{N_S+N_B}$, where $N_S$ ($N_B$) are the number of signal (background) events once the kinematic cuts were applied. We show  in Figs. \ref{significance1}-\ref{significance3} the contour plots of the signal significance as a function of $m_{A_F}$ and the integrated luminosity for the BMP1-BMP3, respectively, as shown in Table \ref{TablaBMP}. 
The results for the case of the $CP$-even Flavon $H_F$, as well as the \texttt{MadGraph} files, will be shown upon request.
 \begin{figure}[!htb]
\centering
\subfigure[ ]{\includegraphics[scale = 0.2]{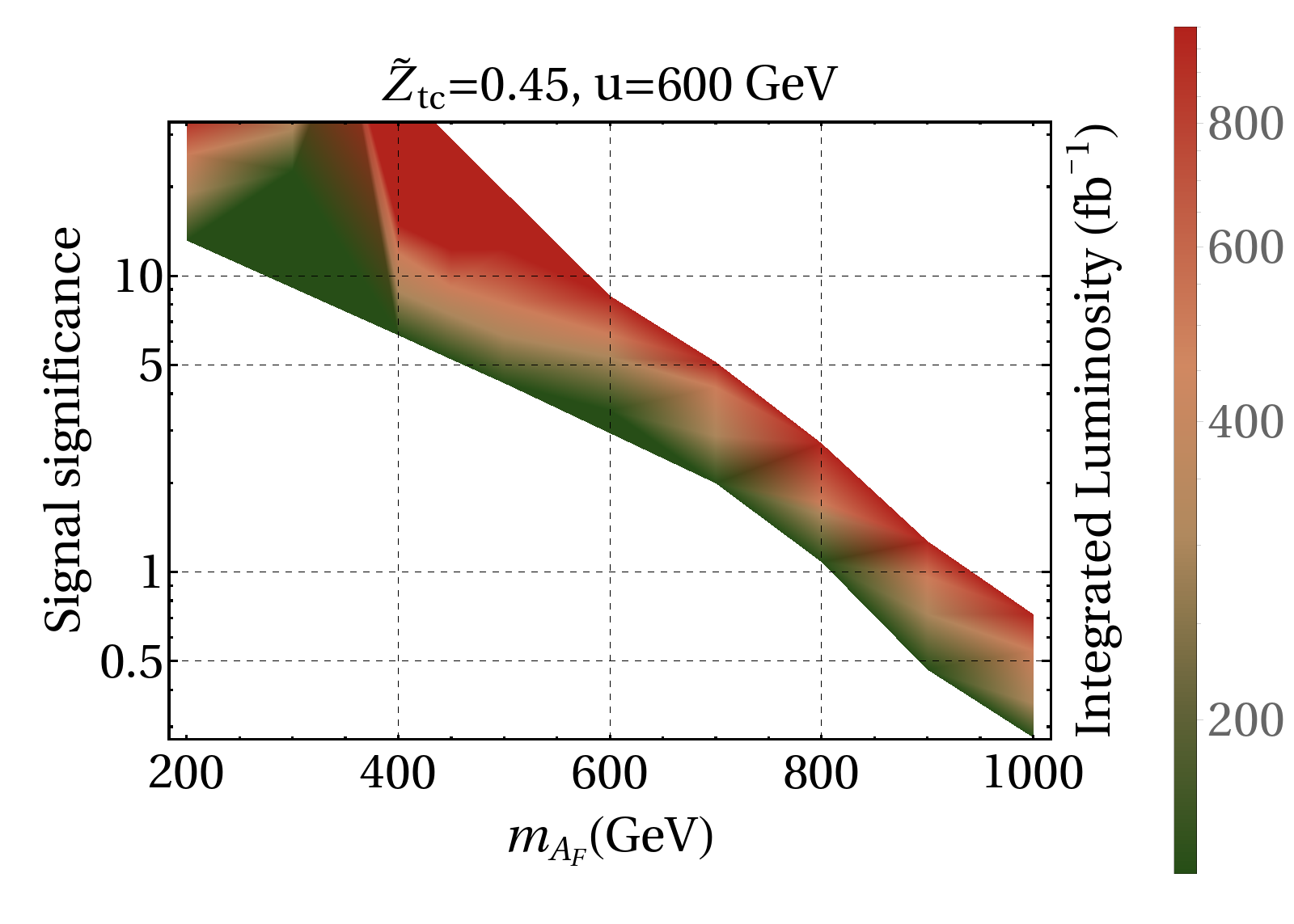}}
\subfigure[ ]{\includegraphics[scale = 0.2]{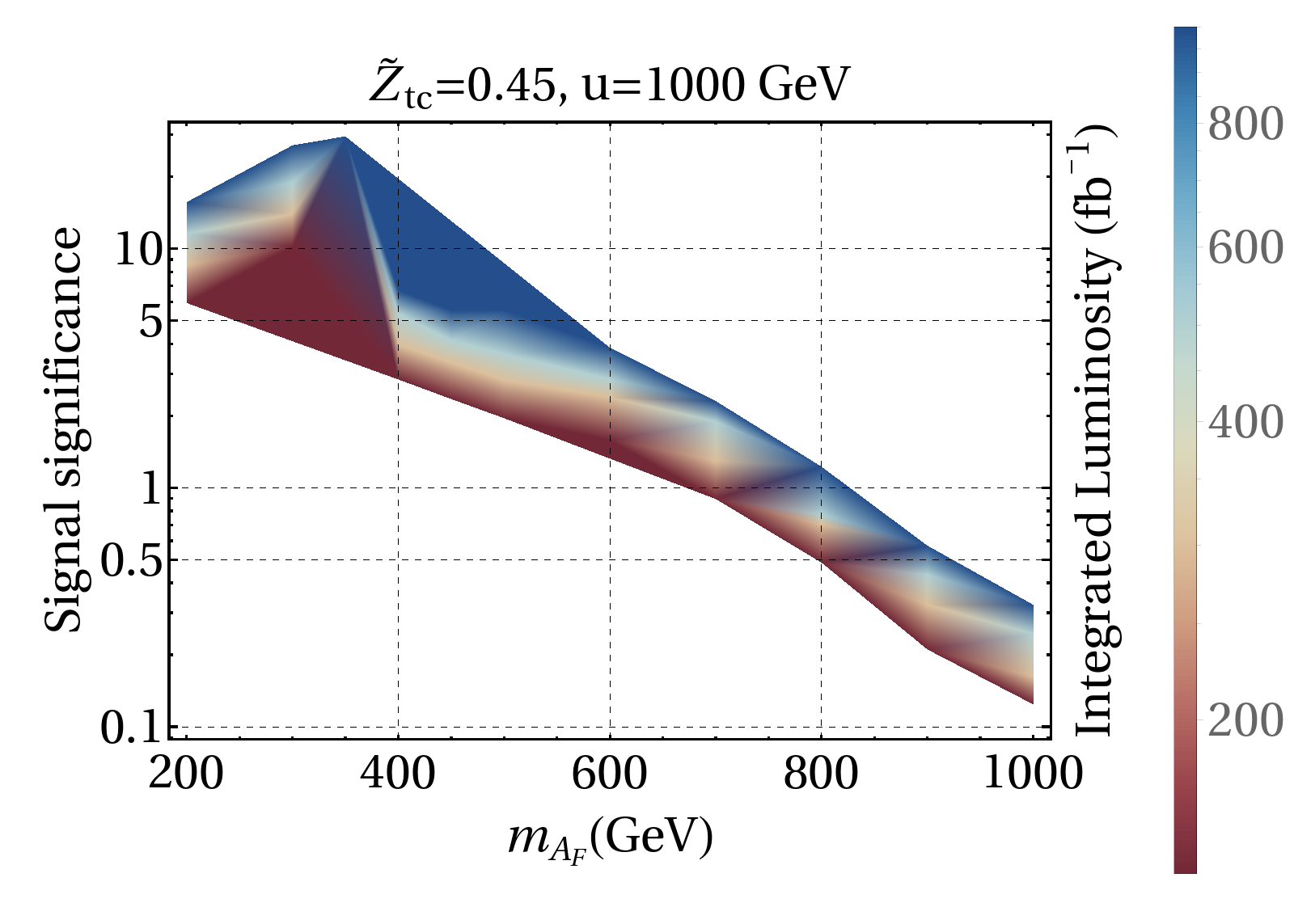}}
       \caption{Contour plots for the signal significance as a function of the integrated luminosity and the $CP-$odd flavon mass, $m_{A_F}$.}  \label{significance1}
	\end{figure}
	
	\begin{figure}[!htb]
\centering
   \subfigure[ ]{\includegraphics[scale = 0.2]{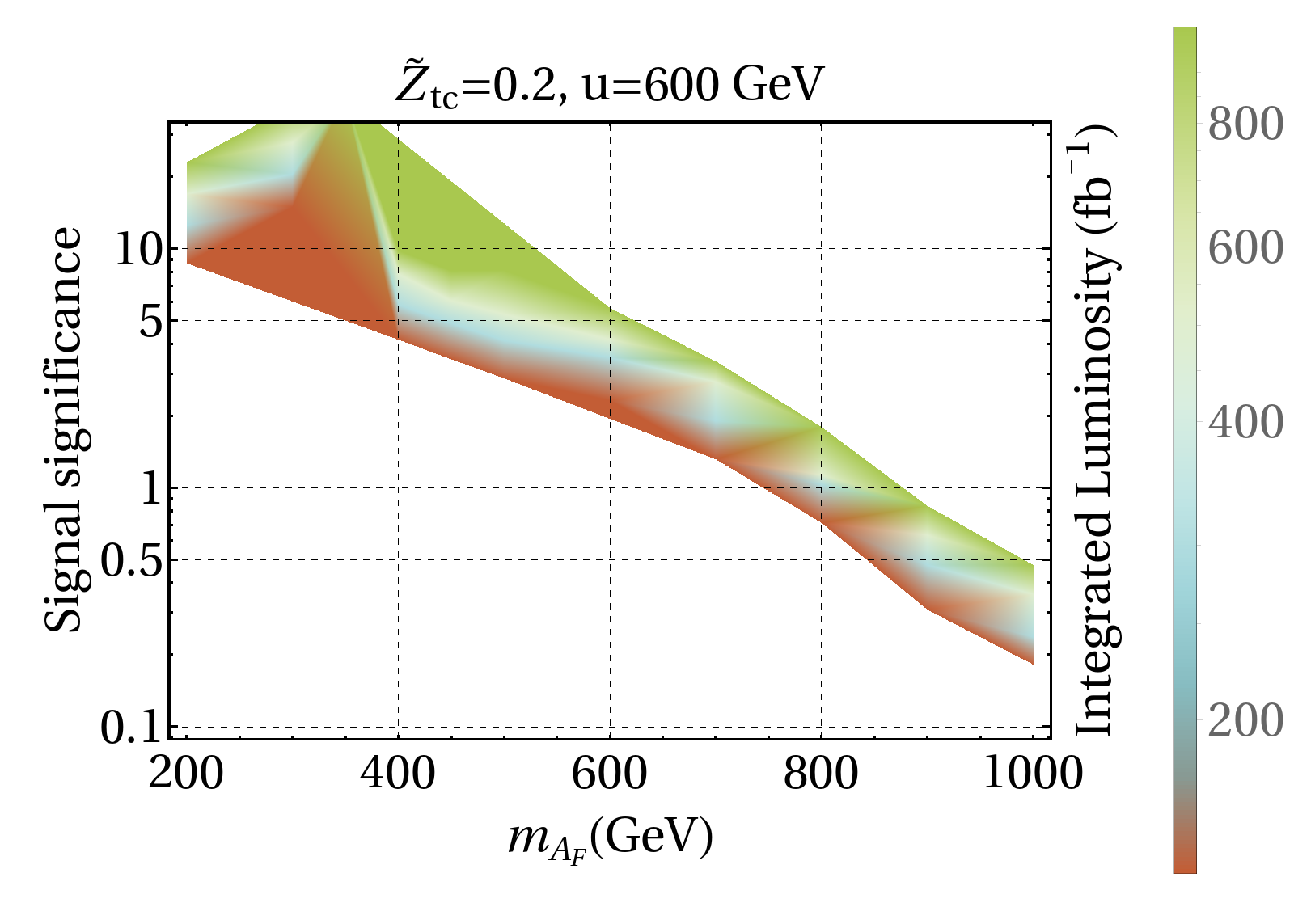}}
   \subfigure[ ]{\includegraphics[scale = 0.2]{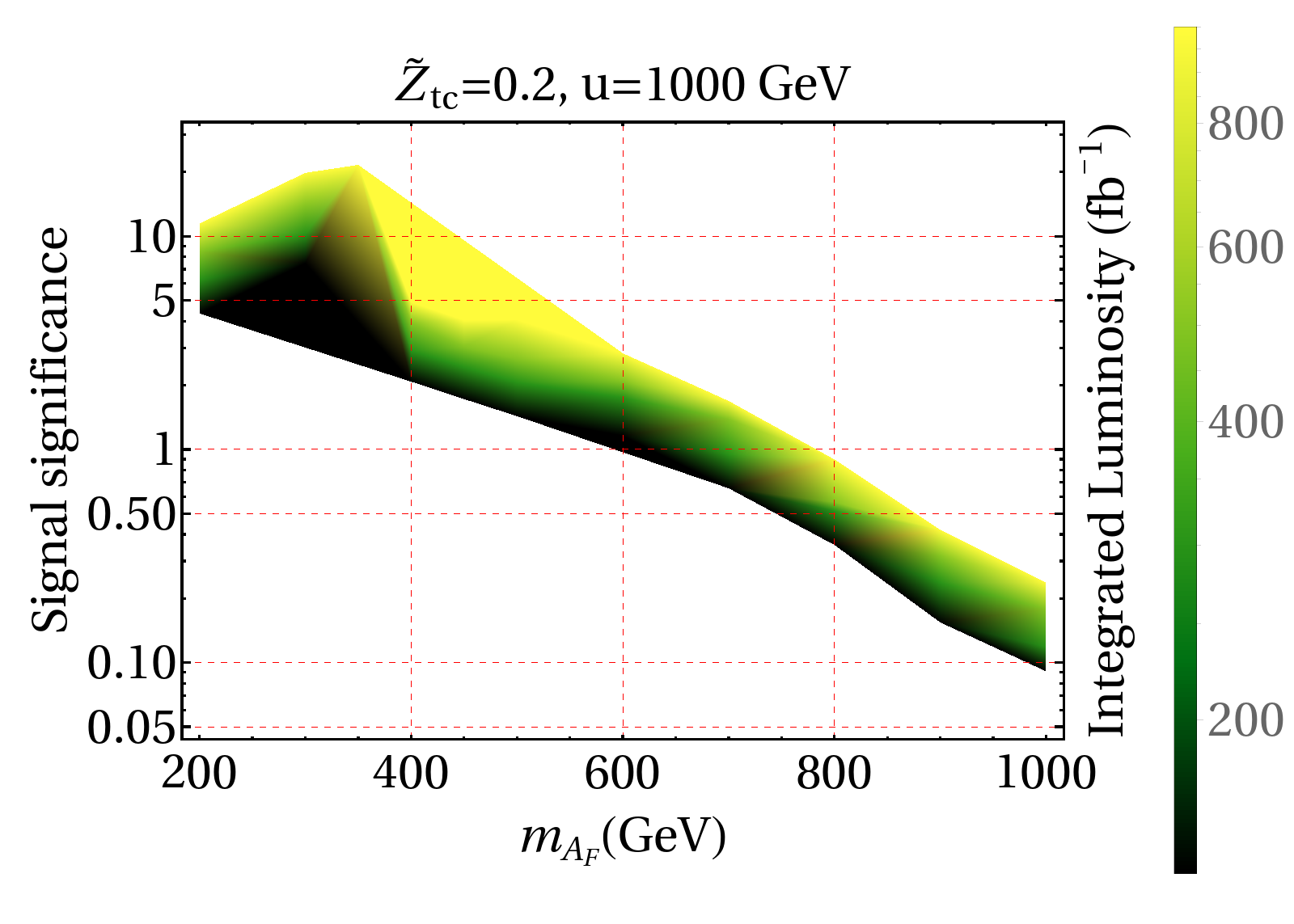}}
    \caption{Contour plots for the signal significance as a function of the integrated luminosity and the $CP-$odd flavon mass, $m_{A_F}$.}  \label{significance2}
	\end{figure}
 
 \begin{figure}[!htb]
\centering
   \subfigure[ ]{\includegraphics[scale = 0.2]{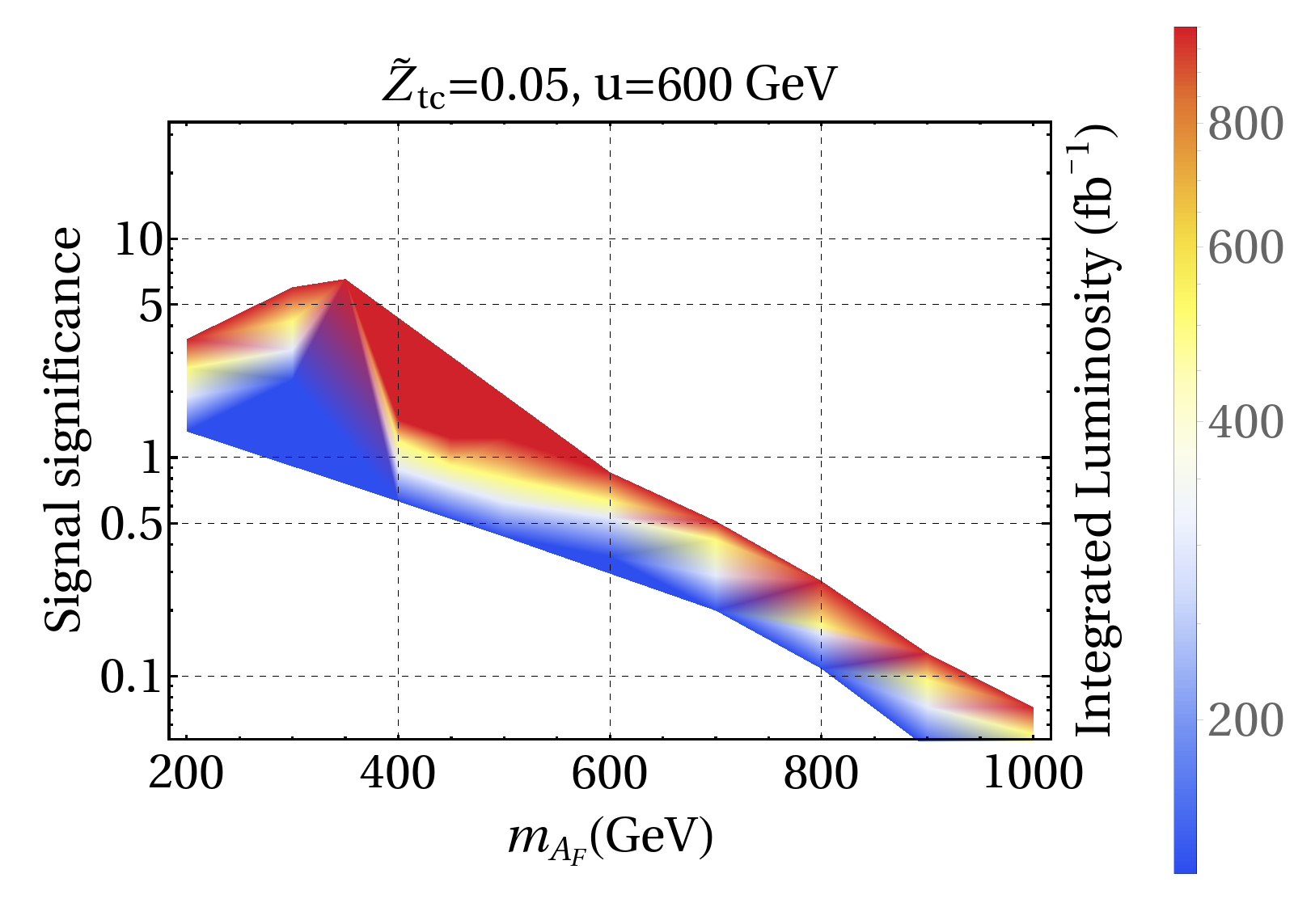}}
      \subfigure[ ]{\includegraphics[scale = 0.2]{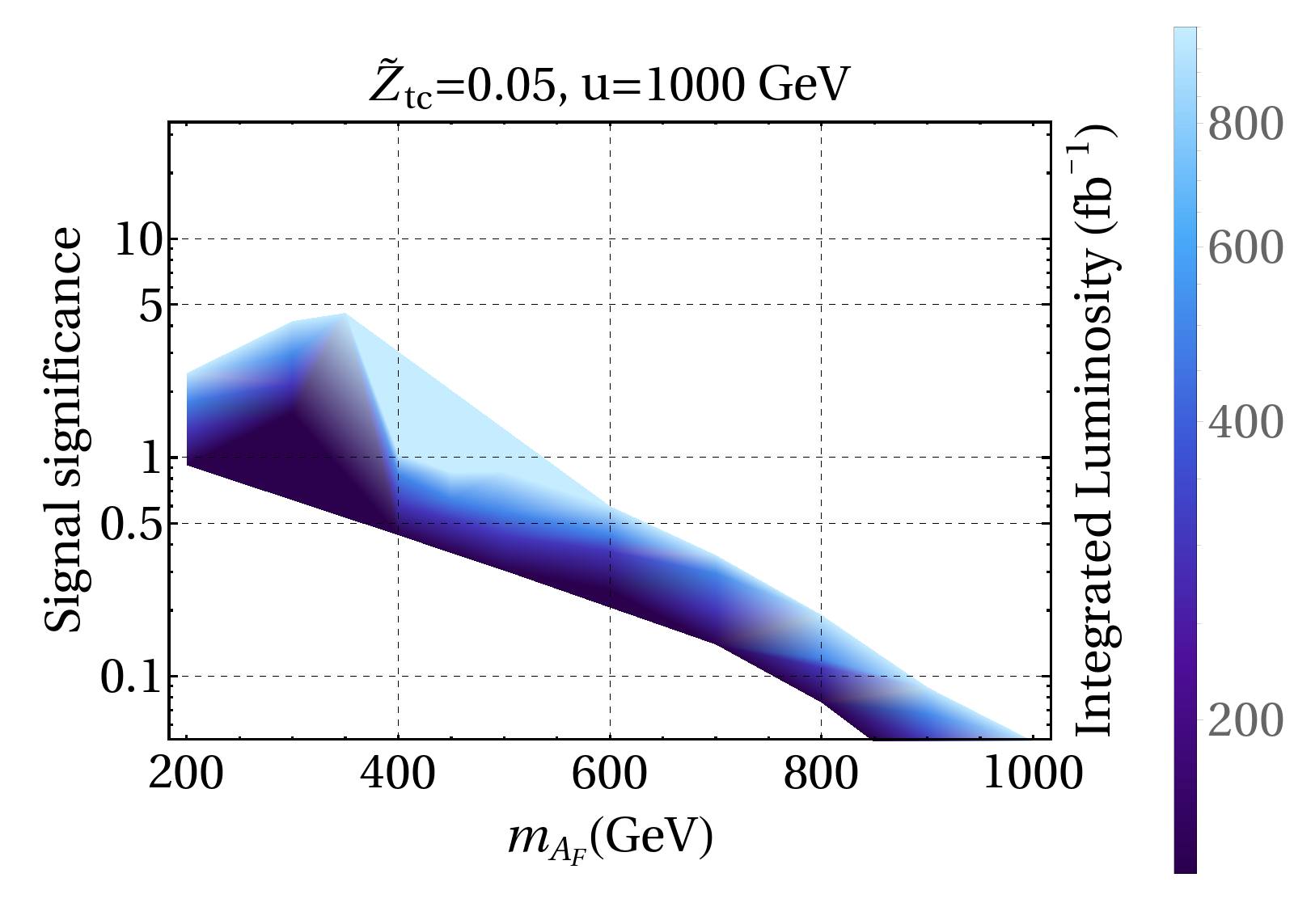}}
    \caption{Contour plots for the signal significance as a function of the integrated luminosity and the $CP-$odd flavon mass, $m_{A_F}$.}  \label{significance3}
	\end{figure}


\newpage
\section{Conclusions\label{sec5}}
We study an extension of the SM with a complex singlet that invokes the Froggatt-Nielsen  mechanism with an Abelian flavor symmetry. Such a model predicts  $CP$-even and  $CP$-odd Flavons that mediate FCNC at tree-level and thus can  decay as $\phi\to tc$ ($\phi=H_F,\,A_F$), which is the focus of our work. We found the region of the parameter space consistent with both experimental and theoretical constraints.  Then, we define a few benchmark points to evaluate the $\phi\to tc$ decays along with the flavon $\phi$ production cross-section at the LHC and its next stage, the HL-LHC. We present a Monte Carlo analysis of both the signal $gg\to \phi\to tc\to b\ell\nu_{\ell}c$ and the main standard model background, focusing on integrated luminosities in the range $140-1000$ fb$^{-1}$, which allow us to assess the possibility that this channel could be detected at the LHC in the best scenario of the model parameters. However, with the advent of the HL-LHC operating to $\mathcal{L}\sim1000$ fb$^{-1}$, it could be possible to detect the decays $\phi\to tc$ for a reasonable scenario in the $200<m_{A_F}<700$ GeV interval and $200<m_{H_F}<380$ GeV. However, if one considers the expected integrated luminosity at the HL-LHC ($3000$ fb$^{-1}$), the mass interval of the Flavons could be increased. We make available, upon request, the necessary files to reproduce the Monte Carlo analysis.

\section*{Acknowledgement}
	We acknowledge support from CONACYT (M\'exico). Partial support from VIEP-BUAP is also acknowledge. The work of M. A. Arroyo-Ure\~na was supported by Centro Interdisciplinario de Investigaci\'on y Ense\~nanza de la Ciencia (CIIEC).

\end{document}